\documentclass[aps,prx,showpacs,amssymb,amsart,superscriptaddress,nofootinbib,twocolumn,longbibliography]{revtex4-1}
\usepackage[a4paper, margin=2.50cm]{geometry}
\usepackage{fancyhdr, color}
\usepackage{tikz}
\usetikzlibrary{shapes,snakes,backgrounds,fit,decorations.pathreplacing}
\usepackage[ansinew]{inputenc}
\usepackage{bbm}
\usepackage{bm}
\usepackage{amsbsy}
\usepackage{amsthm}
\usepackage{amssymb}
\usepackage{amsfonts}
\usepackage{amsmath}
\usepackage{dsfont} % for symbols like the identity: \mathds{1}
\usepackage{graphicx} % for graphics
\usepackage{epsfig}
\usepackage{epstopdf}
\usepackage{enumerate}
\usepackage{dsfont}
\usepackage{multibib}
\usepackage[colorlinks,breaklinks]{hyperref}
\makeatletter
\newcommand\org@hypertarget{}
\let\org@hypertarget\hypertarget
\renewcommand\hypertarget[2]{%
  \Hy@raisedlink{\org@hypertarget{#1}{}}#2%
  }

\makeatother
\usepackage[figure,table]{hypcap}
\usepackage{MnSymbol}
\usepackage{enumerate}%allows different styles of enumerate environment
\usepackage{float}
\hypersetup{
	bookmarksnumbered,
	pdfstartview={FitH},
	citecolor={darkgreen},
	linkcolor={darkred},
	urlcolor={darkblue},
	pdfpagemode={UseOutlines}}
\definecolor{darkgreen}{RGB}{50,190,50}
\definecolor{darkblue}{RGB}{0,0,190}
\definecolor{darkred}{RGB}{238,0,0}
\usepackage{soul}
\newcommand{\Sys}{\ensuremath{_{\hspace{-1pt}\protect\raisebox{0pt}{\tiny{$S$}}}}}
\newcommand{\SA}{\ensuremath{_{\hspace{-1pt}\protect\raisebox{0pt}{\tiny{$A$}}}}}
\newcommand{\supA}{\ensuremath{^{\hspace{-0pt}\protect\raisebox{0pt}{\tiny{$A$}}}}}
\newcommand{\supB}{\ensuremath{^{\hspace{-0pt}\protect\raisebox{0pt}{\tiny{$B$}}}}}
\newcommand{\SB}{\ensuremath{_{\hspace{-1pt}\protect\raisebox{0pt}{\tiny{$B$}}}}}
\newcommand{\R}{\ensuremath{_{\hspace{-1pt}\protect\raisebox{0pt}{\tiny{$R$}}}}}
\newcommand{\SAB}{\ensuremath{_{\hspace{-1pt}\protect\raisebox{0pt}{\tiny{$A\hspace*{-0.5pt}B$}}}}}
\newcommand{\SR}{\ensuremath{_{\hspace{-1pt}\protect\raisebox{0pt}{\tiny{$S\hspace*{-0.5pt}R$}}}}}
\newcommand{\AR}{\ensuremath{_{\hspace{-1pt}\protect\raisebox{0pt}{\tiny{$A\hspace*{-0.5pt}R$}}}}}
\newcommand{\BR}{\ensuremath{_{\hspace{-1pt}\protect\raisebox{0pt}{\tiny{$B\hspace*{-0.5pt}R$}}}}}
\newcommand{\USR}{\ensuremath{U_{\hspace{-1.3pt}\protect\raisebox{0pt}{\tiny{$S\hspace*{-0.7pt}R$}}}}}

\newcommand{\ket}[1]{\ensuremath{\left|\right.\!{#1}\!\left.\right\rangle}}

\newcommand{\bra}[1]{\ensuremath{\left\langle\right.\!{#1}\!\left.\right|}}

\newcommand{\tr}{\textnormal{Tr}}

\newtheorem{question}{Question}%[section]

\usepackage{filecontents}
\fancypagestyle{firstpage}{
\fancyhead[C]{\sc \color[rgb]{0.4,0.2,0.9}{\small{In: Thermodynamics in the Quantum Regime \textemdash\ Fundamental Aspects and New Directions\\
F. Binder, L. A. Correa, C. Gogolin, J. Anders, and G. Adesso (eds.)\\
Chapter 30, pp.~731--750, Springer International Publishing, 2019}
}}
\fancyhead[R]{}
\fancyhead[L]{}
}
\fancypagestyle{otherpages}{
\fancyhead[C]{\sc \color[rgb]{0.4,0.2,0.9}{\small{In: Thermodynamics in the Quantum Regime}
}}
\fancyhead[R]{}
\fancyhead[L]{}
}

\begin{document}

\title{Trade-off Between Work and Correlations in Quantum Thermodynamics}
\author{Giuseppe Vitagliano}
\email{giuseppe.vitagliano@univie.ac.at}
\affiliation{Institute for Quantum Optics and Quantum Information - IQOQI Vienna, Austrian Academy of Sciences, Boltzmanngasse 3, 1090 Vienna, Austria}
\author{Claude Kl{\"o}ckl}
%\thanks{Current address: Institute for Sustainable Economic Development, University of Natural Resources and Life Sciences, Gregor-Mendel-Stra{\ss}e 33, 1180 Vienna, Austria}
\email{claude.kloeckl@boku.ac.at}
\affiliation{Institute for Sustainable Economic Development, {University of Natural}\\ Resources and Life Sciences, Gregor-Mendel-Stra{\ss}e 33, 1180 Vienna, Austria}
\affiliation{Institute for Quantum Optics and Quantum Information - IQOQI Vienna, Austrian Academy of Sciences, Boltzmanngasse 3, 1090 Vienna, Austria}
\author{Marcus Huber}
\email{marcus.huber@univie.ac.at}
\affiliation{Institute for Quantum Optics and Quantum Information - IQOQI Vienna, Austrian Academy of Sciences, Boltzmanngasse 3, 1090 Vienna, Austria}
\author{Nicolai Friis}
\email{nicolai.friis@univie.ac.at}
\affiliation{Institute for Quantum Optics and Quantum Information - IQOQI Vienna, Austrian Academy of Sciences, Boltzmanngasse 3, 1090 Vienna, Austria}

\date{\today}

\begin{abstract}
Quantum thermodynamics and quantum information are two frameworks for employing quantum mechanical systems for practical tasks, exploiting genuine quantum features to obtain advantages with respect to classical implementations. While appearing disconnected at first, the main resources of these frameworks, work and correlations, have a complicated yet interesting relationship that we examine here. We review the role of correlations in quantum thermodynamics, with a particular focus on the conversion of work into correlations. We provide new insights into the fundamental work cost of correlations and the existence of optimally correlating unitaries, and discuss relevant open problems.
\end{abstract}

\maketitle

\thispagestyle{firstpage}

%%%%%%%%%%%%%%%%%%%%%%%%%%%%%%%%%%%%%%%%%%%%%%%%%%%%%%%%%%%%%%%%%%%%%%%%%%%%%%%%%%%%%%%%%%%%%%%%%%%%

\section{Introduction}\label{sec:introduction}
%\vspace*{-1mm}

From its inception, classical thermodynamics has been a practically minded theory, aiming to quantify the usefulness and capabilities of machines in terms of performing work. One of the distinguishing features of quantum thermodynamics with respect to classical thermodynamics (and standard quantum statistics) is the level of control that one assumes to have over the degrees of freedom of microscopic (and mesoscopic) quantum systems. This control allows one to identify genuine quantum behaviour in thermodynamics beyond contributions via statistical deviations, e.g., due to the indistinguishability of bosons and fermions. Moreover, this control permits us to consider individual quantum systems as the figurative gears of a quantum machine \textemdash quantum ``cogwheels" that can be used to extract or store work, which act as catalysts, or mediate interactions. The operating principle of such a (quantum) machine is to use the control over the system to convert one kind of resource into another, for instance, thermal energy into mechanical work. Quantum thermodynamics can hence be viewed as a resource theory~\cite{BrandaoHorodeckiOppenheimRenesSpekkens2013}, where energy\footnote{Or, more generally, out-of-equilibrium states from which energy can be extracted.} is a resource, while systems in thermal equilibrium with the environment are considered to be freely available. The control over the system then determines how energy can be spent to manipulate (thermal) quantum states.\\

At the same time, the manipulation of well-controlled quantum systems is the basic premise for quantum information processing~\cite{NielsenChuang2000}. The latter, in turn, can also be understood as a resource theory with respect to (quantum) correlations~\cite{HorodeckiOppenheim2013a, EltschkaSiewert2014}. However, it is clear that for any practical application, that this abstract information-theoretic construct must be embedded within a physical context, e.g., such as that provided by quantum thermodynamics. It therefore seems natural to wonder, how the resources of quantum thermodynamics and quantum information theory are related, how they can be converted into each other, and what role resources of one theory play in the other, respectively.\\

Here, we want to discuss these questions and review the role of (quantum) correlations in quantum thermodynamics (see Refs.~\cite{GooldHuberRieraDelRioSkrzypczyk2016, MillenXuereb2016, VinjanampathyAnders2016} for recent reviews). In Section~\ref{sec:literature review}, we briefly review a few key areas within quantum thermodynamics where (quantum) correlations are of significance, starting with an overview of the basic concepts and definitions in Sections~\ref{subsec:frameworkI} and \ref{subsec:frameworkII}. In Sections~\ref{sec:Role of correlations for thermodynamic laws} and~\ref{sec:role of correlations for work extraction}, the consequences of the presence of correlations for the formulation of thermodynamic laws and for the task of work extraction are discussed, respectively.\\

In Section~\ref{sec:optimally correlating unitaries and bound work}, we then focus on the specific question of converting the resource of quantum thermodynamics, i.e., energy, into the resource of quantum information, i.e., correlations. To this end, we first review previous key results~\cite{HuberPerarnauHovhannisyanSkrzypczykKloecklBrunnerAcin2015, BruschiPerarnauLlobetFriisHovhannisyanHuber2015, GiorgiCampbell2015, FriisHuberPerarnauLlobet2016, FrancicaGooldPlastinaPaternostro2017} that provide bounds on the performance of this resource conversion, before turning to a, thus far, unresolved problem: the question of the existence of optimally correlating unitaries. We show that such operations do not always exist and analyse the implications of this observation. Finally, we discuss pertinent open problems in understanding the role of correlations in thermodynamics.

\newpage
\pagestyle{otherpages}

\section{Correlations in Quantum Thermodynamics}\label{sec:literature review}
\vspace*{-1mm}
%%%%%%%%%%%%%%%%%%%%%%%%%%%%%%%%%%%%%%%%%%%%%%%%%%%%%%%%%%%%%%%%%%%%%%%%%%%%%%%%%%%%%%%%%%%%%%%%%%%

\subsection{Framework}\label{sec:framework}
\vspace*{-1mm}

\subsubsection{Quantum thermodynamics in a nutshell}
\label{subsec:frameworkI}
\vspace*{-1mm}

In the following, we consider pairs of quantum mechanical systems that may share correlations. In the context of thermodynamics, these might be, e.g., a system (working body) and a heat bath. In quantum mechanics, these systems are encoded into a bipartite Hilbert space $\mathcal{H}\SAB=\mathcal{H}\SA \otimes \mathcal{H}\SB$, where the tensor factors represent the two subsystems, $A$ and $B$. States describing the joint system are given by density operators $\rho\SAB\in\mathcal{L}(\mathcal{H}\SAB)$, i.e., positive-semidefinite ($\rho\SAB\geq0$), linear operators over $\mathcal{H}\SAB$ that satisfy $\tr(\rho\SAB)=1$. The energy of the system is further determined by the Hamiltonian $H\SAB$, i.e., a self-adjoint operator over $\mathcal{H}\SAB$ usually assumed to be bounded from below. Besides the average (internal) energy $E(\rho\SAB)=\tr(H \rho\SAB)$ of the joint system, other state functions central to quantum thermodynamics are the {\it von~Neumann entropy} $S(\rho\SAB)=-\tr\bigl(\rho\SAB\ln(\rho\SAB)\bigr)$ and the {\it free energy} $F(\rho\SAB)=E(\rho\SAB)-TS(\rho\SAB)$, as well as the analogous quantities for the reduced states $\rho\SA=\tr\SB(\rho\SAB)$ and $\rho\SB=\tr\SA(\rho\SAB)$ of the subsystems.

The focus of quantum thermodynamics then lies on the study of the (possible) evolution of $\rho\SAB$ and the subsystem states $\rho\SA=\tr\SB(\rho\SAB)$ and $\rho\SB=\tr\SA(\rho\SAB)$ subject to certain constraints on the allowed transformations of the joint system, such as, for example:
\begin{enumerate}[(i)]
\item{Conservation of (total) energy of $A$ and $B$:
    \begin{align}
        \rho\SAB\mapsto \sigma\SAB: \tr(H\SAB \rho\SAB) = \tr(H\SAB \sigma\SAB)\nonumber
    \end{align}}
\item{Closed system dynamics:
    \begin{align}
        \rho\SAB\mapsto U\SAB \rho\SAB U^\dagger\SAB\nonumber
    \end{align}
    for some global unitary transformation $U\SAB$.}
\end{enumerate}

To state its basic laws, then, a first main goal of quantum thermodynamics is to provide meaningful definitions of {\it heat} $\Delta Q$ and {\it work} $\Delta W$  exchanged between the subsystems. These quantities are not just functions of the state, but depend on the concrete transformations that are applied. In particular, suitable definitions are usually chosen such that the {\it first law of thermodynamics} holds for closed joint systems, i.e.,
\begin{align}
   \Delta E &=\, \Delta Q + \Delta W \,,
   \label{eq:heat and work}
\end{align}
with $\Delta E=E(\rho\SA(t+\Delta t))-E(\rho\SA(t))$. For this purpose a quantifier often used for the information change within the subsystems is the \textit{relative entropy} $S(\rho\SA(t+\Delta t) \| \rho\SA(t))$ of $\rho\SA(t+\Delta t)$ w.r.t. $\rho\SA(t)$. For arbitrary states $\rho$ and $\sigma$ it is defined as
\begin{align}
   S(\sigma \| \rho) &=\, -S(\sigma) - \tr\bigl(\sigma\ln(\rho)\bigr) \,.
\end{align}
The relative entropy is nonnegative, $S(\sigma \| \rho)\geq 0$, for all pairs $\rho$ and $\sigma$, with equality iff $\rho=\sigma$ (see, e.g.,~\cite{Vedral2002} for details), and coincides with the free energy difference $S(\sigma \| \rho)=F(\sigma)-F(\rho)$, whenever $\rho=\tau(\beta)$ is thermal. That is, given a Hamiltonian $H\SAB$, the corresponding {\it thermal state} is defined as
\begin{align}
    \tau\SAB(\beta) &=\,\mathcal{Z}^{-1}e^{-\beta H\SAB}\,,
    \label{eq:thstatedef}
\end{align}
where $\beta=1/T$ denotes the inverse temperature\footnote{We use units where $\hbar=k_{\protect\raisebox{-0pt}{\tiny{B}}}=1$ throughout.} and $\mathcal{Z}=\tr(e^{-\beta H\SAB})$ is called the {\it partition function}. The above thermal state $\tau\SAB(\beta)$ can be considered as the state of the joint system of $A$ and $B$ at thermal equilibrium with an external heat bath at temperature $T$, and represents the maximum entropy state for fixed (internal) energy~\cite{Jaynes1957}. The relative entropy can hence be understood as measure of distance from thermal equilibrium of the total system.

\vspace*{-3mm}
\subsubsection{Quantifying correlations}
\label{subsec:frameworkII}
\vspace*{-3mm}

The properties of equilibrium states strongly depend on the system Hamiltonian. For instance, the joint thermal state  $\tau_{AB}$ of systems $A$ and $B$ is completely uncorrelated whenever they are noninteracting, i.e., when $H\SAB=H\SA+H\SB$, where $H\SA$ and $H\SB$ act nontrivially only on $\mathcal{H}\SA$ and $\mathcal{H}\SB$, respectively. In this case $\tau\SAB$ is a product state $\tau\SAB(\beta)=\tau\SA(\beta)\otimes\tau\SB(\beta)$, where $\tau\SA$ and $\tau\SB$ are thermal states at the same temperature $T=1/\beta$ w.r.t. the local Hamiltonians $H\SA$ and $H\SB$, respectively. Conversely, in an interacting system the global state $\rho\SAB$ will typically  be correlated, and may even be entangled, meaning that the state cannot be decomposed into a mixture of uncorrelated states (see, e.g.,~\cite{FriisHuberPerarnauLlobet2016} for a discussion).

Such correlations can be quantified in terms of the \textit{mutual information}\footnote{This is one of the most widely used measures of correlations in the context of thermodynamics, which arises quite naturally, due to being a linear function of the von Neumann entropies of the subsystems, and thus directly related to thermodynamical potentials. See also the subsequent discussion.}
\begin{align}
    \mathcal{I}(\rho\SAB)   &=S(\rho\SA)+S(\rho\SB)-S(\rho\SAB).
    \label{eq:mutual information}
\end{align}
That is, $\mathcal{I}$ quantifies the amount of information about the system that is available globally but not locally. Here, it is important to note $\mathcal{I}$ captures both classical and genuine quantum correlations in the sense that nonzero values of $\mathcal{I}$ may originate in either type of correlation. Nonetheless, it should be noted that any state $\rho\SAB$ for which $\mathcal{I}(\rho\SAB)>S(\rho\SA)$ or $\mathcal{I}(\rho\SAB)>S(\rho\SB)$ features a negative conditional entropy $S(A\| B)=S(\rho\SA)-\mathcal{I}(\rho\SAB)$, and is hence~\cite{CerfAdami1999, DelRioAbergRennerDahlstenVedral2011} necessarily entangled to some extent\footnote{See, e.g.,~\cite{FriisBulusuBertlmann2017} for a pedagogical introduction to entanglement detection via conditional entropies and mutual information.}.

%%%%%%%%%%%%%%%%%%%%%%%%%%%%%%%%%%%%%%%%%%%%%%%%%%%%%%%%%%%%%%%%%%%%%%%%%%%%%%%%%%%%%%%%%%%%%%%%%%%

\subsubsection{Role of correlations for thermodynamic laws}\label{sec:Role of correlations for thermodynamic laws}

Indeed, correlations are already central to the foundations of quantum thermodynamics. This manifests in the fundamental difference between describing thermodynamic systems as being composed of isolated parts, or as interacting with each other. When subsystems are considered to be completely isolated, just as in classical thermodynamics, this translates to the quantum mechanical notion of a product state $\rho\SA \otimes \rho\SB$. However, there are continuing efforts to expand and even challenge this seemingly basic assumption. This includes, e.g., approaches where subsystems (and hence potential correlations between them) are defined using thermodynamic principles~\cite{StokesDebBeige2017}, or those where work and heat exchange in interacting systems is defined in terms of effective local Hamiltonians that depend on correlations~\cite{WeimerHenrichRemppSchroederMahler2008, TeifelMahler2011}. In other approaches, the traditional setting of a system separable from its surrounding thermal bath can be extended to a state $\rho\SAB$ with \textit{correlations between system and bath}~\cite{Partovi2008, JenningsRudolph2010, JevticJenningsRudolph2012a, JevticJenningsRudolph2012b, BrandaoHorodeckiOppenheimWehner2015, AlipourBenattiBakhshinezhadAfsaryMarcantoniRezakhani2016, NathBeraRieraLewensteinWinter2017, Mueller2018}. This leads to reformulations of the concepts of heat and work and modifications of the classical laws of thermodynamics via the introduction of correlated baths.

In particular, the second law deserves a special mention in this respect. Loosely speaking, it states that the entropy of a subsystem cannot decrease after a thermodynamical transformation, i.e., $\Delta S = S(\rho\SA(t+\Delta t))-S(\rho\SA(t)) \geq 0$, an observation that lies at the heart of the emergence of the so-called {\it thermodynamical arrow of time}. However, several authors, have observed how such a relation does not hold true anymore when system and bath are allowed to be initially correlated \cite{Partovi2008, JenningsRudolph2010}. This insight can even be traced back to Boltzmann himself~\cite{Boltzmann1870, Boltzmann1897}, who, while introducing the so-called {\it Stosszahlansatz} (i.e., the assumption of molecular chaos), noticed that in order for the second law of thermodynamics to emerge it was necessary to assume a weakly correlated (cosmological) environment.

Formally speaking, the basic argument showing the intimate connection of correlations with the thermodynamical arrow of time works as follows in its extremal version~\cite{Partovi2008, JenningsRudolph2010, JevticJenningsRudolph2012a, JevticJenningsRudolph2012b}. Consider a system and bath with local Hamiltonians $H\SA$ and $H\SB$, such that $H\SA\ket{n}\SA=E_{n}\supA\ket{n}\SA$ and $H\SB\ket{n}\SB=E_{n}\supB\ket{n}\SB$. Suppose that there exist constants $\mu\SA$ and $\mu\SB$ such that the local energy levels $E_{n}\supA$ and $E_{n}\supB$ satisfy $\mu\SA E_{n}\supA = \mu\SB E_{n}\supB=\epsilon_n$ for all $n$. Let us further assume that initially system and bath are jointly very close to a (pure) {\it highly entangled} state\footnote{For simplicity here we could consider finite-dimensional systems.}
\begin{align}\label{eq:purebithermal}
   \ket{\psi}\SAB &=\, \mathcal{Z}^{-1/2} \sum_n \exp(-\gamma \epsilon_n/2)  \ket{n,n}\SAB \,,
\end{align}
which can be thought of as a thermal state of a suitable interacting Hamiltonian $H\SAB$ for very low temperatures (i.e., in the limit $\beta \rightarrow \infty$). The two marginal states $\rho\SA$ and $\rho\SB$ are thermal w.r.t. $H\SA$ and $H\SB$ at (different) inverse temperatures $\beta\SA = \mu\SA \gamma$ and $\beta\SB = \mu\SB \gamma$ respectively, but with the same entropy $S(\rho\SA) = S(\rho\SB)$ since $\rho\SA$ and $\rho\SB$ have the same spectrum. Suppose now that the two subsystems interact through a global unitary transformation that allows some exchange of energy between the subsystems. Further, let us assume that, w.r.t. to a suitable definition of $Q$ [see the discussion surrounding Eq.~(\ref{eq:heat and work})], this exchange is interpreted simply as a heat exchange, meaning that the local changes of internal energy satisfy $\Delta E\SA=\Delta Q\SA$ and $\Delta E\SB=\Delta Q\SB$, respectively. Since the marginals are initially thermal, we have $\Delta F\SA=\Delta E\SA-T\SA\Delta S\SA\geq0$ and $\Delta F\SB=\Delta E\SB-T\SB\Delta S\SB\geq0$, which leads to
$\beta\SA \Delta Q\SA\geq \Delta S\SA$ and $\beta\SB \Delta Q\SB\geq \Delta S\SB$, while $\Delta Q\SA+\Delta Q\SB=0$. Globally, the constraint on heat exchanges implies
\begin{align}
  \beta\SA \Delta Q\SA + \beta\SB \Delta Q\SB &\geq \, \Delta \mathcal I\SAB \,,
\end{align}
where $\Delta\mathcal I\SAB=\Delta S\SA + \Delta S\SB$ is the change in the mutual information after the transformation since the unitary leaves the overall entropy invariant. The crucial observation is then that for the above global state $\ket{\psi}\SAB$ the mutual information can be very high initially and can decrease during the transformation such that $\Delta \mathcal I\SAB <0$. Heat may thus be allowed to flow from the cold to the hot subsystem (e.g., $\Delta Q\SA \geq 0$ for $\beta\SA \leq \beta\SB$). Correlations can thus lead to an \emph{anomalous heat flow}, see also%~\chqfs
~\cite{LevyGelbwaserKlimovsky2019}. Moreover, since the above global state remains pure, the marginals have the same spectra, and we have $\Delta S\SA = \Delta S\SB$. Therefore, an anomalous heat flow implies a violation of the classical second law of thermodynamics, $\Delta \mathcal I\SAB <0 \ \Rightarrow \ \Delta S\SA=\Delta S\SB <0$, i.e., the local entropies both decrease, reversing the direction of the thermodynamical arrow of time.

Following this observation, several authors, adopting an information theoretic perspective, have investigated the possibility to generalize the thermodynamic laws in the presence of initial correlations between system and bath, see, e.g., Refs.~\cite{DelRioHutterRennerWehner2016, BrandaoHorodeckiOppenheimWehner2015, AlipourBenattiBakhshinezhadAfsaryMarcantoniRezakhani2016, NathBeraRieraLewensteinWinter2017, Mueller2018}. Furthermore, experiments are now being performed for quantum mechanical systems realized in several platforms to observe violations of classical laws of thermodynamics, especially regarding the inversion of the thermodynamic arrow of time \cite{MicadeiEtAl2017}.

Correlations therefore need to be carefully incorporated into the formulation of quantum thermodynamics. However, correlations are not only a source of seemingly paradoxical situations but can also have direct practical relevance for paradigmatic tasks in quantum thermodynamics, as we will discuss in Section~\ref{sec:role of correlations for work extraction}, before we analyse what quantum thermodynamics tells us about the work cost and level of control necessary to create correlations in Section~\ref{sec:optimally correlating unitaries and bound work}.

%%%%%%%%%%%%%%%%%%%%%%%%%%%%%%%%%%%%%%%%%%%%%%%%%%%%%%%%%%%%%%%%%%%%%%%%%%%%%%%%%%%%%%%%%%%%%%%%%%%

\subsection{Extracting work from correlations}\label{sec:role of correlations for work extraction}

\subsubsection{Work extraction using cyclic transformations}

A basic but crucial application of thermodynamics is to quantify how much work can be extracted from a given machine operating under a {\it cycle of transformations}. In this context, one may consider subsystem $A$ to be a quantum machine that is controlled by the external subsystem $B$. As a general model of such a machine one usually considers an ensemble of, say, $N$ quantum mechanical units, i.e., subsystem $A$ has a Hilbert space $\mathcal H\SA=\mathcal H^{\otimes N}$ with a given (free) Hamiltonian $H\SA$. This allows statements about the scaling of the quantum machine's efficiency and eventual gain (e.g., originating from the ability to create correlations) with $N$ as compared to analogous classical machines.

The external control is usually modelled as a switchable time-dependent (and cyclic) Hamiltonian $H(t)$, such that $H(0)=H(t_{\rm cycle})$ where $t_{\rm cycle}$ is the time of a whole cycle. The following question then arises naturally: \emph{Can correlations between the $N$ units of the machine help in extracting work during a thermodynamical cycle?}

More specifically, if we evolve the initial state $\rho\SA$ with an externally controlled unitary transformation $U$ and compare the resulting difference in energy we get the quantity
\begin{equation}
    \Delta W_U=\tr(\rho\SA H\SA) - \tr(U \rho\SA U^{\dag} H\SA).
    \label{eq: energy difference}
\end{equation}
If positive, it describes the amount of energy that is gained after $U$ and is usually interpreted as extracted work, i.e., the process is assumed to
be performed adiabatically with the external control \cite{PerarnauLlobetHovhannisyanHuberSkrzypczykBrunnerAcin2015}.
Thus, frequently (\ref{eq: energy difference}) is seen as a figure of merit that should be maximized with respect to the available resources, like the initial state $\rho\SA$, the externally controlled evolution $U$ and the free system Hamiltonian $H\SA$. Fixing or optimizing over the triple ($\rho\SA, U, H\SA$) allows one to ask questions like: \emph{Which combination of resources yields the most work?} A frequently studied special case is that of \emph{ergotropy}~\cite{AllahverdyanBalianNieuwenhuizen2004}, corresponding to a fixed Hamiltonian and a fixed state while optimizing over all unitaries on $\mathcal{H}\SA$, i.e.,
\begin{equation}
    \Delta W_{\rm ergotropy}:= \max_{U} \ \Delta W_U.
    \label{eq: ergotropy}
\end{equation}
States that do not allow for work extraction with respect to a specified class of operations (typically unitary transformations) are called {\it passive}~\cite{PuszWoronowicz1978}.

In other words, for any state $\rho$ the unitary realizing the maximum in Eq.~(\ref{eq: ergotropy}) is the one that transforms the state to a corresponding passive state $U\rho\, U^\dagger = \rho_{\rm passive}$, and the ergotropy represents the work that is extractable from the system with the specified operations. Quantum systems in passive states thus have a simple practical interpretation as the analogues of \textit{empty batteries}.

%%%%%%%%%%%%%%%%%%%%%%%%%%%%%%%%%%%%%%%%%%%%%%%%%%%%%%%%%%%%%%%%%%%%%%%%%%%%%%%%%%%%%%%%%%%%%%%%%%%

\subsubsection{Role of correlations for work extraction}

An interesting distinction between the concepts of passive and thermal states that has recently been discovered~\cite{AlickiFannes2013} is the following. A state is passive if and only if it is diagonal in the energy eigenbasis and its eigenvalues are decreasing with increasing energy. While this is certainly true for thermal states, it can also be the case for many other eigenvalue distributions which are not thermal. In summary, thermality implies passivity, but the converse is not necessarily true.

The special role of thermal states comes to light when considering many copies of the system: While $\tau(\beta)^{\otimes N}$ is still thermal (and thus passive) for any $N$, $\rho_{\mathrm{passive}}^{\otimes N}$ is passive for all $N$ if and only if $\rho_{\mathrm{passive}}$ is a thermal state~\cite{Lenard1978, AlickiFannes2013}. In other words, thermal states are the only {\it completely passive} states. This has interesting consequences when interpreting passive states as empty batteries. While a single battery appears empty, i.e., no work whatsoever can be extracted, it may be the case that adding a second empty battery would enable a correlating global transformation to extract work out of the two empty batteries. This interesting situation is termed \textit{work extraction by activation}.

This fact leads to a first affirmative answer to the question of whether correlating operations can help for work extraction: Unitary transformations can extract work more efficiently if they are able to generate entanglement. More precisely, a quantitative analysis of the relation between entanglement generation and work extraction~\cite{HovhannisyanPerarnau-LlobetHuberAcin2013} shows that the trade-off is more accurately specified as occurring between entangling power and the number of required operations (in other words, time) for work extraction: The less entanglement is generated during the work extraction process, the more time is needed to extract work. However, note that the final state need not be entangled.

The above example of two empty batteries showing the difference between local passivity and (true global) passivity can be employed to construct another enlightening example.
In analogy to the batteries above, let us define a state $\rho\SAB$ that is locally thermal (thus locally passive) in each marginal, i.e., $\tr\SB(\rho\SAB)=\tau\SA(\beta)$ and $\tr\SA(\rho\SAB)=\tau\SB(\beta)$. Let us now imagine that, contrary to the previous example, the state $\rho\SAB$ is correlated. If we can extract work from this state, then it can be argued that all the work must have come from its correlations, since there can be no contributions from local operations (due to the passivity of the marginals) nor from activation, due to the fact that thermal states do not allow for work extraction by activation. An example of such a state is provided in~\cite{PerarnauLlobetHovhannisyanHuberSkrzypczykBrunnerAcin2015}. For locally thermal states of noninteracting systems all correlations hence imply extractable work.

%%%%%%%%%%%%%%%%%%%%%%%%%%%%%%%%%%%%%%%%%%%%%%%%%%%%%%%%%%%%%%%%%%%%%%%%%%%%%%%%%%%%%%%%%%%%%%%%%%%

\subsubsection{Role of correlations for work storage}\label{sec:correlations_vs_workstorage}

A problem that can be considered dual to the above is how to efficiently charge an initially empty quantum battery (see also%~\chqbat
~\cite{CampaioliPollockVinjanampathy2019}). Formally, the problem is to find a suitable way of transforming a state $\rho$ to another state $\sigma=U\rho U^\dagger$ such that the latter contains more extractable work, i.e., $\Delta W_U \leq 0$ as in Eq.~(\ref{eq: energy difference}). As previously for work extraction, one is primarily interested in charging processes based on cycles of transformations. However, in contrast to the previous problem, one is here not necessarily interested in asking how much work may be stored in principle. Instead other figures of merit become important, indicating certain desirable properties of the charging process or the final state for fixed $\Delta W_U$. Examples for such properties include charging power~\cite{BinderVinjanampathyModiGoold2015}, fluctuations during the charging process or the variance of the final charge~\cite{FriisHuber2018}. The key question that we wish to discuss here is: \emph{Is it beneficial for work storage to (be able to) generate correlations during the process?}\\

While correlations themselves turn out not to be directly relevant in any crucial way, it appears that the control over correlating transformations can provide significant advantages, even if no actual correlations are created. Specifically, in Ref.~\cite{BinderVinjanampathyModiGoold2015} the authors show that the power of charging a battery, defined as $P:=\langle \Delta W_U \rangle /\Delta t$, i.e., the ratio of average work $\langle \Delta W_U \rangle $ and time $\Delta t$, can be enhanced by allowing entangling global unitaries $U$ on an initially uncorrelated product state $\rho_1 \otimes \dots \otimes \rho_N$. However, whether such entangling operations actually create entanglement during the cycle is irrelevant~\cite{CampaioliPollockBinderCeleriGooldVinjanampathyModi2017}. As is discussed in more detail in%~\chqbat
~\cite{CampaioliPollockVinjanampathy2019}, the trade-off is rather between entanglement generated during the process and the speed of the process itself. Focussing on the practical aspects of implementing this idea, in Ref.~\cite{FerraroCampisiAndolinaPellegriniPolini2018} the authors study a non-unitary charging process of a quantum battery of $N$ qubits coupled to a single photonic mode.

%%%%%%%%%%%%%%%%%%%%%%%%%%%%%%%%%%%%%%%%%%%%%%%%%%%%%%%%%%%%%%%%%%%%%%%%%%%%%%%%%%%%%%%%%%%%%%%%%%%

\subsubsection{Work extraction and storage with restricted control}

For understanding fundamental bounds on work extraction and storage correlations are thus of significance, or rather, the ability to create them. This highlights the fact that this ability is linked to the control one has over the system and operations thereon. In particular, maximization such as in Eq.~(\ref{eq: ergotropy}) may yield solutions that cannot be practically implemented or whose realization comes at a high cost itself. Therefore, subsequent works have focused on understanding the limitations of work extraction and work storage in terms of more restricted sets of states/operations such as Gaussian states and unitaries~\cite{BrownFriisHuber2016, FriisHuber2018}. This has been motivated also by the easier practical implementation in CV systems such as encountered in quantum optics of Gaussian operations, as opposed to arbitrary unitaries. In particular, this is manifest when considering driven transformations, where Gaussian operations appear as the simplest type of operation according to the hierarchy of driving Hamiltonians, which are at most quadratic in the system's annihilation and creation operators for Gaussian unitaries.

As we have already emphasized, passivity is defined with respect to an underlying class of state transformations, i.e., unitaries (cyclic Hamiltonian processes). An interesting variant of the problem in Eq.~(\ref{eq: ergotropy}) is thus given by the restricted case of Gaussian unitary transformations~\cite{BrownFriisHuber2016}. This leads to the notion of \emph{Gaussian passivity} as a special case of passivity in the sense that passivity implies Gaussian passivity, but not vice versa. The role of correlations in such a scenario is particularly interesting. In particular, the authors show that it is always possible to extract work from an entangled Gaussian state via two-mode squeezing operations. For general non-Gaussian states, however, the ability to extract work via (dis)entangling Gaussian unitaries (two-mode squeezing) does not indicate entanglement, and the inability to do so does not imply separability either.

In the complementary problem of battery charging, the subset of Gaussian operations turns out to provide a trade-off between good precision (energy variance) and practical implementability of a battery charging protocol with a fixed target amount of energy $\Delta W$~\cite{FriisHuber2018}. Here, correlations can provide minor advantages in some cases but do not play a conceptually important role.

To conclude this section it is also interesting to point out that, as observed in Ref.~\cite{BrunelliGenoniBarbieriPaternostro2017} the argument that entanglement (or the ability to create it) can provide an advantage in work extraction can also be turned around and exploited to design entanglement certification schemes based on extractable work. In other words, it is possible to witness entanglement by quantifying the extracted work from a thermodynamic cycle, such as the {\it Szilard engine} considered in~\cite{BrunelliGenoniBarbieriPaternostro2017}. Interestingly, the entanglement criterion based on work extraction becomes necessary and sufficient for two-mode Gaussian states. However, the authors also show that the above scheme cannot be applied for the certification of genuine multipartite entanglement.

%%%%%%%%%%%%%%%%%%%%%%%%%%%%%%%%%%%%%%%%%%%%%%%%%%%%%%%%%%%%%%%%%%%%%%%%%%%%%%%%%%%%%%%%%%%%%%%%%%%%
%%%%%%%%%%%%%%%%%%%%%%%%%%%%%%%%%%%%%%%%%%%%%%%%%%%%%%%%%%%%%%%%%%%%%%%%%%%%%%%%%%%%%%%%%%%%%%%%%%%%

\section{Energy Cost of Creating Correlations}\label{sec:optimally correlating unitaries and bound work}

\subsection{Trade-off between work and correlations}\label{sec:trade off starting section}

Now that we have an overview of the importance of correlations (and the transformations that can create them) for quantum thermodynamics and its paradigmatic tasks, let us consider the situation from a different perspective. Correlations, in particular, entanglement, are the backbone of quantum information processing. However, if an abstract information theory is to be applied in practice, it requires a physical context, such as is provided by quantum thermodynamics. There, as we have seen above, the freely available equilibrium state of two (noninteracting) systems is uncorrelated. This means that both an investment of energy and a certain level of control (the ability to perform correlating transformations) are required to create the desired correlations.

Thus, formally speaking, one is interested in determining the fundamental limits for the energy cost of creating correlations, where we choose to quantify the latter by the mutual information of Eq.~(\ref{eq:mutual information}) between two quantum systems $A$ and $B$, since the mutual information arises quite naturally in the thermodynamic context. These subsystems are assumed to be initially in a joint thermal state $\tau\SAB(\beta)=\tau\SA(\beta)\otimes\tau\SB(\beta)$ with respect to a noninteracting Hamiltonian $H\SAB=H\SA+H\SB$ at a particular ambient temperature $T=1/\beta$. To correlate $A$ and $B$, it is necessary to move the joint system out of equilibrium, which comes at a nonzero work cost $W$. For instance, when one acts unitarily on the system, this cost can be expressed as $W=\Delta E\SA + \Delta E\SB$. The question at hand is then: \emph{What is the maximal amount of correlation between $A$ and $B$ that can be reached, given a fixed amount of available energy $W$?}

Clearly, the answer very much depends on the operations that are allowed, as well as on the energy level structure of the local Hamiltonians. For instance, one may consider applying only global unitary transformations $U\SAB$, realized via some external control. If one wishes to optimally convert work into correlations, it is clear that the marginals of the final state must be passive. Otherwise, energy extractable by local unitaries (which leave $\mathcal{I}\SAB$ invariant) would be left in the system. Indeed, the marginals must be completely passive (thus thermal), such that no work can be extracted locally\footnote{Here, local refers to the collections of subsystems $A_{1}, \ldots, A_{N}$ and $B_{1}, \ldots, B_{N}$ for $N$ copies of $\rho\SAB$.} from any number of copies.
The optimally correlated target state $\rho\SAB=U\SAB \tau\SAB(\beta) U\SAB^\dagger$ must hence be such that $\rho\SA=\tr\SB(\rho\SAB)=\tau\SA(\beta\SA)$ and $\rho\SB=\tr\SA(\rho\SAB)=\tau\SB(\beta\SB)$. At the same time, one wishes to increase the correlations, i.e., to achieve the maximal amount of mutual information increase $\Delta \mathcal{I}\SAB=\Delta S\SA + \Delta S\SB\geq 0$. At fixed average energy input $\Delta E=E(\tau\SA(\beta\SA))+E(\tau\SB(\beta\SB))-E(\tau\SAB(\beta))$, the mutual information is then maximized for maximal $S(\tau\SA(\beta\SA))+S(\tau\SB(\beta\SB))=S(\tau\SA(\beta\SA)\otimes\tau\SB(\beta\SB))$. The maximum entropy principle then suggests that the final state marginals should be thermal at the same temperature $\beta\SA=\beta\SB$. But do unitaries exist that can achieve this?

The above requirements are indeed already quite strong, as we can observe through the following extremal example. Let us imagine that initially the two subsystems are in a pure (ground) state (i.e., the limit $\beta\rightarrow \infty$) $\tau\SAB(\beta)=\ket{0}\!\!\bra{0}\SA \otimes \ket{0}\!\!\bra{0}\SB$ at zero initial energy. Then, any unitary will still output a pure state $\ket{\psi}\SAB=U\SAB\ket{00}\SAB$, and the requirement of the marginals to be thermal states leads to a state as in Eq.~(\ref{eq:purebithermal}), i.e.,
\begin{align}
   \ket{\psi}\SAB &=\, \mathcal{Z}^{-1/2} \sum_n \lambda_n \ket{n,n}\SAB \,,
\end{align}
with $\lambda_n^{2}=\exp(-\beta\SA E\supA_{n})=\exp(-\beta\SB E\supB_{n})$. The requirement of having a fixed amount of energy available further demands
\begin{align}
   W   &=\sum_{n} \lambda_{n}^{2} (E\supA_{n}+E\supB_{n}),
\end{align}
which in general translates into a complicated relation between the two final effective inverse temperatures $\beta\SA$ and $\beta\SB$.

Thus, further constraints, e.g., on the final temperatures of the marginal states, on the initial Hamiltonians of the two systems, or on the external control available might already lead to an impossibility of achieving such an optimal conversion between work and correlations. In the following we will make this statement more precise, discuss the relevant questions arising in this context, and give some (partial) answers.

\subsection{Fundamental cost of correlations}\label{sec:fund cost of correlations}

With the realization that there is a finite work cost for the creation of correlations in noninteracting systems~\cite{HuberPerarnauHovhannisyanSkrzypczykKloecklBrunnerAcin2015, BruschiPerarnauLlobetFriisHovhannisyanHuber2015}, or for the increase of correlations in the presence of correlated thermal states for interacting systems~\cite{FriisHuberPerarnauLlobet2016}, two immediate pertinent questions for the trade-off between the resources work and correlations can be formulated. (i) On the one hand, it is of interest to understand the fundamental limitations on achievable correlations without any restrictions on the complexity of the involved operations or the time these may take. We can thus ask: \emph{What is the theoretical minimum work cost for any amount of correlation between two given systems?} (ii) On the other hand, it is of course of practical importance to learn what can be achieved under practical conditions, i.e., in finite time and with limited control over the system, that is: \emph{What is the minimum work cost of correlations that is practically achievable?}

Let us formulate those questions more precisely following the treatment in~\cite{BruschiPerarnauLlobetFriisHovhannisyanHuber2015}.
We first observe that any physical transformation of a system $S$ can be thought of as a unitary map $\USR$ acting on a larger Hilbert space that includes an external reservoir $R$. In this context one may further assume that $S$ and $R$ are initially not correlated, i.e., that initially we have $\tau\SR(\beta)=\tau\Sys(\beta)\otimes\tau\R(\beta)$. Then, the work cost to bring such a state out of equilibrium is
\begin{align}
    W   &=\tr\Bigl(H\bigl[\rho\SR-\tau\SR(\beta)\bigr]\Bigr)=\Delta E\Sys + \Delta E\R,
    \vspace*{-2mm}
    \label{eq:W in}
\end{align}
where the overall final state is $\rho\SR=\USR\hspace*{0.5pt}\tau\SR(\beta)\USR^{\dagger}$, and if we further assume that $H=H\Sys+H\R$, then the energy contributions split up into $\Delta E\Sys=\tr\bigl(H\Sys [\rho\SR-\tau\SR(\beta)]\bigr)$ and $\Delta E\R=\tr\bigl(H\R [\rho\SR-\tau\SR(\beta)]\bigr)$. We note also that similar expressions in related contexts can be found, e.g., in~\cite{OppenheimHorodeckiMPR2002, OppenheimHorodeckiKMPR2003, EspositoVanDenBroeck2011, ReebWolf2014}. One may then rewrite Eq.~(\ref{eq:W in}) by expressing the internal energy differences via the changes in free energy and entropy and obtain
\begin{align}
    W &=\,\Delta F\Sys+\Delta F\R\,+\,T\,\mathcal{I}\SR\, ,
    \label{eq:free energy difference S vs B detailed proof}
\end{align}
where one makes also use of the fact that the initial thermal state is uncorrelated, $S(\tau\SR)=S(\tau\Sys)+S(\tau\R)$ and that $U\SR$ leaves the global entropy invariant, i.e., $S(\rho\SR)=S(\tau\SR)$.

In the above expression, one recognizes the mutual information $\mathcal{I}\SR$ between the system and the reservoir. In complete analogy to the reasoning that leads to Eq.~(\ref{eq:free energy difference S vs B detailed proof}), one may further rewrite the free energy change of the system $S$ comprising the subsystems $A$ and $B$ as
\begin{align}
    \Delta F\Sys    &=\,\Delta F\SA+\Delta F\SB\,+\,T\,\mathcal{I}\SAB\,.
    \label{eq:free energy diff system}
\end{align}
Combining Eqs.~(\ref{eq:free energy difference S vs B detailed proof}) and~(\ref{eq:free energy diff system}) and noting that the initial states of $A$, $B$ and $R$ are thermal, such that the free energy differences can be expressed via the relative entropy, one thus arrives at
\begin{align}
    \beta W   &=\,S(\rho\R \| \tau\R)+S(\rho\SA \| \tau\SA)+S(\rho\SB \| \tau\SB)+\mathcal{I}\SR+\mathcal{I}\SAB,
    \label{eq:work cost of correlations all terms}
\end{align}
where $\rho\SA=\tr\BR(\rho\SR)$,  $\rho\SB=\tr\AR(\rho\SR)$, and $\rho\R=\tr\SAB(\rho\SR)$ are the reduced states after the transformation. Since $S(\rho\|\sigma)\geq0$ for all $\rho$ and $\sigma$ and $\mathcal{I}\SR\geq0$ as well, it becomes clear that the fundamental upper bound for the correlations between $A$ and $B$ is
\begin{align}
    \mathcal{I}\SAB &\leq\beta W.
    \label{eq:fundamental bound for correlation cost}
\end{align}

\begin{figure}[ht!]
\label{fig:protocol}%(Color online)
%%%trim={<left> <lower> <right> <upper>}
\includegraphics[width=0.47\textwidth,trim={0cm 0mm 0cm 0mm}]{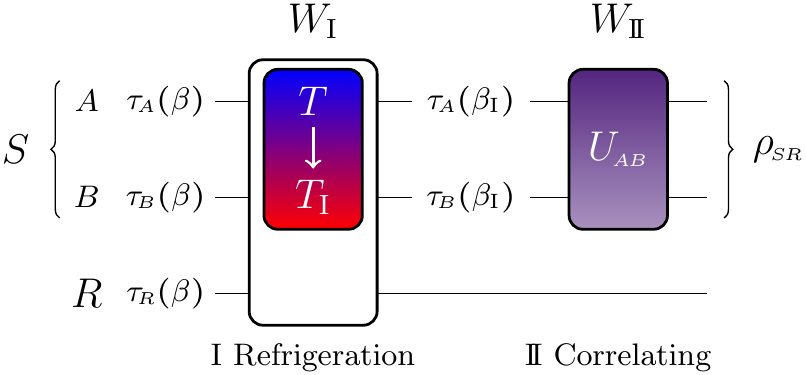}
\vspace*{-4.5mm}
\caption{
\textbf{Two-step correlating protocol:} The temperature of the initially uncorrelated subsystems $A$ and $B$ forming the system $S$ is first lowered from $T$ to $T_{\mathrm{I}}<T$ using the interaction with the reservoir $R$ at the expense of the work $W_{\mathrm{I}}$. In a second step, the reservoir is decoupled from the system, before $A$ and $B$ are correlated by a unitary $U\SAB$, at energy cost $W_{\mathrm{I\!I}}$.}
\end{figure}

It is then crucial to note that tightness of this bound is only given when $S(\rho\R \| \tau\R)=0$ and $\mathcal{I}\SR=0$. On the one hand, these conditions are trivially met when one performs unitary operations acting solely on the system $S$, but not on $R$. We will return to this scenario in Section~\ref{sec: opt corr unitaries}. On the other hand, one could assume perfect control over an arbitrarily large reservoir $R$, i.e., one that is complex enough to thermalize the system whenever $S$ and $R$ come in contact~\cite{Aberg2013} (see also~\cite{SkrzypczykShortPopescu2014} for description of the involved unitaries), such that $R$ can be assumed to be left in its original state with no correlations created between $S$ and $R$.

A protocol that operates based on the latter premise was presented in~\cite{BruschiPerarnauLlobetFriisHovhannisyanHuber2015} and consists of two steps to reach the optimal trade-off between work and correlations, i.e., $\mathcal{I}\SAB=\beta W$, see Fig.~\ref{fig:protocol}. In the first step with work cost $W_{\mathrm{I}}$, the temperature of the system $S$ is lowered from $T$ to $T_{\mathrm{I}}=1/\beta_{\mathrm{I}}<T$. Here one uses the contact to $R$ to refrigerate $S$, resorting to arbitrarily slow processes to do so, such that the cooling cost is given by the free energy difference, $W_{\mathrm{I}}=F\bigl(\tau\Sys(\beta_{\mathrm{I}})\bigr)-F\bigl(\tau\Sys(\beta)\bigr)$. In the second step, a unitary $U\SAB$ acting only on $S$ is used to correlate $A$ and $B$ at work cost $W_{\mathrm{I\hspace*{-0.5pt}I}}$. It is assumed that $U\SAB$ is of such a form that the marginals are returned to locally thermal states at the original temperature $T$. That is, they have to satisfy $\rho\SA=\tau\SA(\beta)$ and $\rho\SB=\tau\SB(\beta)$ in order to obtain $S(\rho\SA \| \tau\SA)=S(\rho\SB \| \tau\SB)=0$ in Eq.~(\ref{eq:work cost of correlations all terms}). This last requirement determines the splitting of the work cost $W$ into $W_{\mathrm{I}}$ and $W_{\mathrm{I\hspace*{-0.5pt}I}}$. One thus finds that there exists a (low-energy) regime, at least in principle, with a linear trade-off between work and correlations, provided that one can exert the mentioned rigorous control over the degrees of freedom of $R$ and that the desired optimal unitaries $U\SAB$ exist.

However, even if this is so, there is a threshold input energy, above which the conversion can only occur sublinearly~\cite{BruschiPerarnauLlobetFriisHovhannisyanHuber2015}. That is, when $\beta W>S\bigl(\tau\Sys(\beta)\bigr)$, the conditions above mean that the ground state is reached in the first step of the protocol, and the excess energy is invested into the second step (a situation similar to the example mentioned in the above preliminary discussion). While the protocol is still optimal, one nonetheless has the strict inequality $\mathcal{I}\SAB<\beta W$.

\subsection{Optimally correlating unitaries}\label{sec: opt corr unitaries}

However, open questions remain associated to the protocol discussed in the previous section. First, the protocol makes use of the conjectured existence of the unitaries $U\SAB$, allowing to reach a final state with $\rho\SA=\tau\SA(\beta)$ and $\rho\SB=\tau\SB(\beta)$ starting from $\tau\Sys(\beta_{\mathrm{I}})$, i.e., with final state marginals that are both effectively at the original temperature. Second, one may question the practicality of the assumptions about the control over $R$ and whether $W_{\mathrm{I}}=F\bigl(\tau\Sys(\beta_{\mathrm{I}})\bigr)-F\bigl(\tau\Sys(\beta)\bigr)$, $S(\rho\R \| \tau\R)=0$ and $\mathcal{I}\SR=0$ may be achieved within reasonable (time) constraints. Both of these issues connect to the previously mentioned scenario where one operates exclusively (and unitarily) on the closed systems $S$. There, the control requirement on $R$ is relaxed from assuming the ability to perform arbitrary unitaries on the overall Hilbert space, to that of isolating and unitarily acting on the significantly smaller system $S$.

Moreover, both situations raise the question whether the respective optimal unitaries exist. In the case of the unitaries $U\SAB$ for step $\mathrm{I\hspace*{-0.5pt}I}$ of the protocol described in the previous section, the special requirement to reach $\mathcal{I}\SAB=\beta W$ in the low-energy regime where $\beta W\leq S\bigl(\tau\Sys(\beta)\bigr)$, is that the effective temperatures of the marginals of the final state are both the original temperature $T=1/\beta$. Mathematically, we can phrase the question of the existence of $U\SAB$ like this:

\begin{question}\label{question 1}
Does there exist a unitary $U\SAB$ on $\mathcal{H}\SAB$ such that
\begin{align}
    \rho\SA &=\tr\SB\bigl(U\SAB\tau\SAB(\beta_{\mathrm{I}}) U\SAB^{\dagger}\bigr)=\tau\SA(\beta),\\
    \rho\SB &=\tr\SA\bigl(U\SAB\tau\SAB(\beta_{\mathrm{I}}) U\SAB^{\dagger}\bigr)=\tau\SB(\beta),
\end{align}
for every pair of local Hamiltonians $H\SA$ and $H\SB$, for all temperatures $T_{\mathrm{I}}=1/\beta_{\mathrm{I}}$ (after the step~$\mathrm{I}$) and all initial (and thus effective final) temperatures $T=1/\beta\geq T_{\mathrm{I}}$?
\end{question}

For some important special cases, Question~\ref{question 1} can be answered affirmatively. For instance, it was shown in~\cite{HuberPerarnauHovhannisyanSkrzypczykKloecklBrunnerAcin2015} that such optimally correlating unitaries exist whenever the local Hamiltonians are identical, $H\SA=H\SB$, and either all energy levels are equally spaced, i.e., and $E_{m+1}-E_{m}=E_{n+1}-E_{n}$ for all $m,n$ such that $H\SA\ket{n}=E_{n}\ket{n}$, or for arbitrary spacings when the difference between $T_{\mathrm{I}}$ and $T$ is large enough (for a quantitative statement see the appendix of~\cite{HuberPerarnauHovhannisyanSkrzypczykKloecklBrunnerAcin2015}).

In particular, for two qubits, $\mathcal{H}\SA=\mathcal{H}\SB=\mathbb{C}^{2}$ this means that optimal generation of correlations in the low-energy regime is always possible as long as $H\SA=H\SB$, since qubits only posses a single energy gap. However, as the preliminary extremal example treated above in Section~\ref{sec:trade off starting section} suggests, whenever $H\SA \neq H\SB$
the optimally correlating unitaries $U\SAB$ cannot always lead to final states with marginals at the same temperatures, in particular, not in the limiting case when $T_{\mathrm{I}}=0$. To be more precise, the subadditivity of the von Neumann entropy imposes the constraint
\begin{align}
    |S(\tau\SA(\beta)) - S(\tau\SB(\beta))| &\leq S(\tau\SAB(\beta))\,=\,S(\tau\SAB(\beta_{\mathrm{I}}))\nonumber\\
    &=\,S(\tau\SA(\beta_{\mathrm{I}})) + S(\tau\SB(\beta_{\mathrm{I}}))
\end{align}
between the initial and final marginal entropies. This constraint is not automatically satisfied if the energy levels of the Hamiltonians $H\SA$ and $H\SB$ are not equal, since $S(\tau\SA(\beta)) \neq S(\tau\SB(\beta))$ in that case. Let us now also work out a counterexample for $T_{\mathrm{I}}>0$.

Consider a bipartite system with local Hamiltonians $H\SA=\omega\SA\ket{1}\!\!\bra{1}\SA$ and $H\SB=\omega\SB\ket{1}\!\!\bra{1}\SB$ with gaps $\omega\SA$ and $\omega\SB$, respectively, where we have set the ground state energy levels to zero without loss of generality. Let us further define $a_{\mathrm{I}}:=1/\bigl(1+e^{-\beta_{\mathrm{I}}\omega\SA}\bigr)$ and $b_{\mathrm{I}}:=1/\bigl(1+e^{-\beta_{\mathrm{I}}\omega\SB}\bigr)$. The initial state $\tau\Sys(\beta_{\mathrm{I}})$ is then of the form
\begin{small}
\begin{align}
    \tau\Sys(\beta_{\mathrm{I}}) &=
    \operatorname{diag}\{a_{\mathrm{I}}b_{\mathrm{I}},a_{\mathrm{I}}(1\hspace*{-1pt}-\hspace*{-1pt}b_{\mathrm{I}}),
    (1\hspace*{-1pt}-\hspace*{-1pt}a_{\mathrm{I}})b_{\mathrm{I}},(1\hspace*{-1pt}-\hspace*{-1pt}a_{\mathrm{I}})(1\hspace*{-1pt}-\hspace*{-1pt}b_{\mathrm{I}})\}.
    \label{eq:initial state diagonal}
\end{align}
\end{small}
If a unitary $U\SAB$ exists satisfying the requirements of Question~\ref{question 1}, then the local reduced states $\rho\SA$ and $\rho\SB$ must be thermal (at the same temperature $T=1/\beta$), and hence diagonal w.r.t. the respective energy eigenbases. In particular, in the limiting case $\omega\SB\rightarrow\infty$ it is easy to see that the entropies of the single-qubit initial and final marginals for subsystem $B$ are $S(\tau\SB(\beta))=S(\tau\SB(\beta_{\mathrm{I}}))=0$ and the subadditivity constraint above would require $S(\tau\SA(\beta))\leq S(\tau\SA(\beta_{\mathrm{I}}))$, which cannot be satisfied for $\beta<\beta_{\mathrm{I}}$ for finite and nonzero $\omega\SA$. To illustrate this more explicitly, we consider an example for finite temperatures and energy gaps,which for simplicity of presentation assumes only a restricted class of unitaries.  That is, let us assume that only unitaries can be performed such that the density operator of the two-qubit final state $\rho_S = U \tau_S U^\dagger$ is of the form
\begin{align}
    \rho\Sys    &=\begin{pmatrix}
        \rho_{00}   & 0 & 0 & d_{2} \\
        0   & \rho_{01} & d_{1} & 0 \\
        0   & d_{1}^{*} & \rho_{10} & 0 \\
        d_{2}^{*}  & 0 & 0 & \rho_{11}
    \end{pmatrix}\,
\end{align}
for some appropriate $d_{1}, d_{2}\in\mathbb{C}$ and probabilities $\rho_{00}+\rho_{01}+\rho_{10}+\rho_{11}=1$. The corresponding marginals are thus
\begin{align}
    \rho\SA &=\begin{pmatrix}
        \rho_{00}+\rho_{01} & 0 \\
        0 & \!\!\!\!\!\rho_{10}+\rho_{11}
    \end{pmatrix},\
    \rho\SB =\begin{pmatrix}
        \rho_{00}+\rho_{10} & 0 \\
        0 & \!\!\!\!\!\rho_{01}+\rho_{11}
    \end{pmatrix},
\end{align}
such that $\rho_{00}+\rho_{01}=1/\bigl(1+e^{-\beta\omega\SA}\bigr)=:a$ and $\rho_{00}+\rho_{10}=1/\bigl(1+e^{-\beta\omega\SB}\bigr)=:b$. Using these two conditions along with the normalization condition, we can express the diagonal elements of $\rho\Sys$ as $\rho_{01}=a-\rho_{00}$, $\rho_{10}=b-\rho_{00}$, and $\rho_{11}=1+\rho_{00}-a-b$. We can then calculate the eigenvalues $\lambda_{i}$ (for $i=1,2,3,4$) of $\rho\Sys$ in terms of the variables $\rho_{00}$, $a$, $b$, $d_{1}$ and $d_{2}$, obtaining
\begin{subequations}
\begin{align}
    \lambda_{1,4}   &=\rho_{00}+\tfrac{1}{2}(1-a-b)
    \pm\sqrt{\bigl(\tfrac{1-a-b}{2}\bigr)^{2}+|d_{2}|^{2}},\\
    \lambda_{2,3}   &=\tfrac{1}{2}(a+b)-\rho_{00}
    \pm\sqrt{\bigl(\tfrac{a-b}{2}\bigr)^{2}+|d_{1}|^{2}}.
\end{align}
\end{subequations}
For any fixed choice of $\omega\SA$, $\omega\SB$, $\beta$ and $\beta<\beta_{\mathrm{I}}$, the state $\rho\Sys$ lies in the unitary orbit of $\tau\Sys(\beta_{\mathrm{I}})$, when there exist valid choices of $\rho_{00}$, $d_{1}$ and $d_{2}$, such that the ordered list of the $\lambda_{i}$ matches the diagonal entries of $\tau\Sys$ given by Eq.~(\ref{eq:initial state diagonal}). Here, note that since $\lambda_{4}\leq\lambda_{1}$ and $\lambda_{3}\leq\lambda_{2}$, there are in principle $6$ possible ways in which the $\lambda_{i}$ could be ordered. For each of these $6$ combinations, one can express $\rho_{00}$ by adding $\lambda_{4}$ and $\lambda_{1}$ (or, equivalently, $\lambda_{3}$ and $\lambda_{2}$), which eliminates dependencies on the off-diagonals $d_{j}$. For instance, when $\omega\SA=3\omega\SB=3\sqrt{\ln2}$ and $\beta_{\mathrm{I}}=2\beta=2\sqrt{\ln2}$, one finds
\begin{align}
    \tau\Sys(\beta_{\mathrm{I}}) &=\,\operatorname{diag}\{\tfrac{256}{325}, \tfrac{64}{325}, \tfrac{4}{325}, \tfrac{1}{325}\}.
    \label{eq:initial state diagonal example}
\end{align}
and the six possible ways to match with the $\lambda_{i}$ result in the values
\begin{align}
    \rho_{00}   &=\,\tfrac{1}{5850}\times\begin{cases}
    4505 &\text{if}\ \lambda_{4}<\lambda_{1}<\lambda_{3}<\lambda_{2}\\
    3965 &\text{if}\ \lambda_{4}<\lambda_{3}<\lambda_{1}<\lambda_{2}\\
    2237 &\text{if}\ \lambda_{4}<\lambda_{3}<\lambda_{2}<\lambda_{1}\\
    1670 &\text{if}\ \lambda_{3}<\lambda_{2}<\lambda_{4}<\lambda_{1}\\
    2210 &\text{if}\ \lambda_{3}<\lambda_{4}<\lambda_{2}<\lambda_{1}\\
    3938 &\text{if}\ \lambda_{3}<\lambda_{4}<\lambda_{1}<\lambda_{2}
    \end{cases}\,.
    \label{eq:six cases}
\end{align}
At the same time, the positivity of $\lambda_{3}\geq0$ and $\lambda_{4}\geq0$ then demands that
\begin{align}
\bigl(\tfrac{7}{9}-\rho_{00}\bigr)^{2}-\tfrac{1}{81}    &\geq |d_{1}|^{2}\geq0,\\[1mm]
\bigl(\rho_{00}-\tfrac{5}{18}\bigr)^{2}-\bigl(\tfrac{5}{18}\bigr)^{2}    &\geq |d_{2}|^{2}\geq0,
\end{align}
which can be turned into the inequality
\begin{align}
\tfrac{3250}{5850}\leq \rho_{00}\leq\tfrac{3900}{5850}.
\end{align}
Since none of the values in Eq.~(\ref{eq:six cases}) satisfy this inequality,
there are some choices of $\omega\SA$, $\omega\SB$, $\beta_{\mathrm{I}}$ and $\beta<\beta_{\mathrm{I}}$ such that it is impossible to find the corresponding $\rho_{00}$.
Together, these examples illustrate that unitaries allowing to achieve $\mathcal{I}\SAB=\beta W$ in the low-energy regime of the two-step protocol of~\cite{BruschiPerarnauLlobetFriisHovhannisyanHuber2015} do not exist in general.\\
\vspace*{-1mm}

Although the general answer to Question~\ref{question 1} is thus negative, this leaves us with a number of interesting open problems, with which we conclude.

%%%%%%%%%%%%%%%%%%%%%%%%%%%%%%%%%%%%%%%%%%%%%%%%%%%%%%%%%%%%%%%%%%%%%%%%%%%%%%%%%%%%%%%%%%%%%%%%%%%%

\section{Open problems \& conclusion}\label{sec:discussion}

First, we note that a more restricted version of Question~\ref{question 1} for the large class of situations when the local Hamiltonians are identical but not equally spaced (beyond local dimension $2$) remains unanswered\footnote{Since the completion of this review chapter, progress has been made on this problem, including proofs for the existence of optimally correlating unitaries in local dimensions $d=3$ and $d=4$, see Ref.~\cite{BakhshinezhadEtAl2019}}. Moreover, one could ask more generally about the optimal conversion of average energy to mutual information in the unitary case and whether it is possible to find cases, where some of the invested work necessarily gets stuck in locally passive, but not thermal states. Second, while we have seen from Eq.~(\ref{eq:work cost of correlations all terms}) that reaching $\mathcal{I}\SAB=\beta W$ is in general not possible for solely unitary correlating protocols (since this would require $S(\rho\SA \| \tau\SA)=S(\rho\SB \| \tau\SB)=0$ and hence $\rho\SA=\tau\SA$ and $\rho\SB=\tau\SB$), one may ask what the optimal trade-off between work and correlations is in such cases. This question also applies in equal manner to the high-energy regime where $W>S\bigl(\tau\Sys(\beta)\bigr)$, which corresponds exactly to the example in Section~\ref{sec:trade off starting section}.\\
\vspace*{-0.5mm}

Another interesting problem is the question whether it is possible that there is a combination of $H\SA$, $H\SB$, and $\beta$ such that for some given amount of work $W$, the restricted unitary $U\SAB$ required for the low-energy regime of the two-step protocol of Section~\ref{sec:fund cost of correlations} exists, allowing correlations to be created in the amount of $\beta W$, while no optimally correlating unitaries exists that could achieve $\beta W$ directly (without cooling). This would imply that correlations could be created optimally only by using control over the reservoir $R$ for cooling, but the corresponding work value stored in the correlations~\cite{PerarnauLlobetHovhannisyanHuberSkrzypczykBrunnerAcin2015} could not be retrieved unitarily from the system. In this sense work would be \emph{bound} in the system.\\
\vspace*{-0.5mm}

Let us also remark that the protocols and optimal unitaries discussed here apply for the creation of correlations as measured by the mutual information. However, when one restricts to genuine quantum correlations, i.e., entanglement, the situation becomes vastly more complicated, starting with the fact that there are many inequivalent measures and it is in general hard to even calculate how much entanglement is present w.r.t. any of these. Consequently, some simple cases of optimal protocols are known~\cite{HuberPerarnauHovhannisyanSkrzypczykKloecklBrunnerAcin2015, BruschiPerarnauLlobetFriisHovhannisyanHuber2015}, suggesting that, indeed, quite different protocols are required, but much is yet to be discovered. Nonetheless, a measure-independent question is of course the (partial)-separability of quantum states. There is a minimum work cost to turn a thermal product state into a (multipartite) entangled state. In~\cite{HuberPerarnauHovhannisyanSkrzypczykKloecklBrunnerAcin2015}, limiting temperatures for entanglement generation were shown to scale linearly with the number of parties, i.e., $T_{\rm max}\leq O(E \cdot(n-1))$ for an exponentially small work cost, i.e., $W\leq O(E\cdot n \cdot c^{-n})$, where $c$ is a constant. Alternatively, one might also consider correlation quantifiers that do not distinguish between classical and quantum correlations at all, which is of relevance, e.g., when assessing the correlations between a measured system and the measurement apparatus after non-ideal measurement procedures~\cite{GuryanovaFriisHuber2018}.\\
\vspace*{-0.5mm}

\newpage
In conclusion, the interplay of work and correlations provides a fascinating but complex interface between quantum thermodynamics and quantum information theory that reveals interesting quantum effects in thermodynamics and allows for advantages for certain paradigmatic tasks such as work extraction. While some of the questions arising from the conversion of these resources have been addressed, a number of subtle but challenging open problems remain.\\

%%%%%%%%%%%%%%%%%%%%%%%%%%%%%%%%%%%%%%%%%%%%%%%%%%%%%%%%%%%%%%%%%%%%%%%%%%%%%%%%%%%%%%%%%%%%%%%%%%

\bigskip

%%%%%%%%%%%%%%%%%%%%%%%%%%%%%%%%%%%%%%%%%%%%%%%%%%%%%%%%%%%%%%%%%%%%%%%%%%%%%%%%%%%%%%%%%%%%%%%%%%%%

\acknowledgements
\vspace*{-3.5mm}
We are grateful to Faraj Bakhshinezhad, Felix Binder, and Felix Pollock for helpful comments and suggestions. We thank Rick Sanchez for moral support. We acknowledge support from the Austrian Science Fund (FWF) through the  START project Y879-N27, the Lise-Meitner project M 2462-N27, the project P 31339-N27, and the joint Czech-Austrian project MultiQUEST (I 3053-N27 and GF17-33780L).

%%%%%%%%%%%%%%%%%%%%%%%%%%%%%%%%%%%%%%%%%%%%%%%%%%%%%%%%%%%%%%%%%%%%%%%%%%%%%%%%%%%%%%%%%%%%%%%%%%%%

\bibliography{bibfile}

%%%%%%%%%%%%%%%%%%%%%%%%%%%%%%%%%%%%%%%%%%%%%%%%%%%%%%%%%%%%%%%%%%%%%%%%%%%%%%%%%%%%%%%%%%%%%%%%%%%%

\end{document}